\documentclass[conference]{IEEEtran}

\usepackage{cite} 
\usepackage[numbers]{natbib} 
\usepackage{xspace} 
\usepackage{textcomp} 
\usepackage{xifthen} 
\usepackage{url} 
\usepackage{amssymb} 
\usepackage{stmaryrd} 
\usepackage{mathtools} 
\usepackage[inline]{enumitem} 
\usepackage{algorithm} 
\usepackage[noend]{algpseudocode} 
\usepackage[group-separator={,}]{siunitx} 
\usepackage{xcolor} 
\usepackage[hidelinks]{hyperref} 
\usepackage{breakurl}
\usepackage[all]{nowidow} 
\usepackage[keeplastbox]{flushend} 
\usepackage{booktabs} 

\usepackage{pgfplots}
\pgfplotsset{compat=newest}
\usepackage{tikz}
\usetikzlibrary{
    calc,
    arrows.meta,
    decorations.markings, 
    decorations.pathreplacing, 
    matrix, 
}

\usepackage{isabelle,isabellesym}
\isabellestyle{it}

\definecolor{light-gray}{gray}{0.9}
\definecolor{gray}{gray}{0.7}
\definecolor{dark-gray}{gray}{0.5}
\definecolor{darker-gray}{gray}{0.3}

\definecolor{color1}{RGB}{255, 65, 68} 
\definecolor{color1b}{RGB}{252, 91, 99} 

\definecolor{color2}{RGB}{234, 127, 43} 
\definecolor{color2b}{RGB}{253, 174, 97} 

\definecolor{color3}{RGB}{200, 200, 0} 
\definecolor{color3b}{RGB}{255, 255, 191} 

\definecolor{color4}{RGB}{98, 181, 86} 
\definecolor{color4b}{RGB}{171, 221, 164} 

\definecolor{color5}{RGB}{43, 131, 186} 
\definecolor{color5b}{RGB}{158, 204, 239} 

\tikzset{
    hbrace/.style={
        decorate,
        decoration={brace,amplitude=6pt},
        xshift=0pt,
        yshift=-2pt
    },
    hbrace-tiny/.style={
        decorate,
        decoration={brace,amplitude=3pt},
        xshift=0pt,
        yshift=-4pt
    },
}

\tikzset{
    my mark/.style 2 args={
        draw=#1, #1, thick,
        every mark/.append style={fill=#1},
        mark=#2, mark options={scale=0.8}
    }
}
\pgfplotscreateplotcyclelist{custom}{
    my mark={color4}{*},
    my mark={color1}{square},
    my mark={color5}{+},
}
\pgfplotsset{
    cycle list name=custom,
    /pgfplots/error bars/error bar style={thick},
    /pgfplots/error bars/error mark options={thick, mark size=3pt,rotate=90},
}

\ifdefined\SuppressAnnotations
    \newcommand{\annot}[4]{}
    \newcommand{\todo}[2]{}
\else
    \newcommand{\annot}[4]{(\textcolor{#1}{\textbf{#2\protect\ifthenelse{\isempty{#3}}{}{ (#3)}\ifthenelse{\isempty{#4}}{}{:}}} \emph{#4})}
    \newcommand{\todo}[2][]{\annot{color1}{TODO}{#1}{#2}}
\fi

\newcommand{\protocol}[0]{Logres\xspace}

\newcommand{\PhaseA}[0]{Collection\xspace}
\newcommand{\PhaseB}[0]{Consensus\xspace}
\newcommand{\PhaseC}[0]{Signing\xspace}

\newcommand{\algMain}[0]{\textsc{\protocol}\xspace}
\newcommand{\algSub}[0]{\textsc{\PhaseB}\xspace}
\newcommand{\prefixSubProps}[0]{C}

\newcommand{\phaseA}[0]{\expandafter\lowercase\expandafter{\PhaseA}}
\newcommand{\phaseB}[0]{\expandafter\lowercase\expandafter{\PhaseB}}
\newcommand{\phaseC}[0]{\expandafter\lowercase\expandafter{\PhaseC}}

\newcommand{\fMkLog}[0]{\text{makeLog}} \newcommand{\fMkDigest}[0]{\text{getDigest}}

\newcommand{\BFT}[0]{Byzantine fault tolerant\xspace}

\newcommand{\propAgreement}{All valid logs created during a run of the protocol must be equal.}
\newcommand{\propCompleteness}{%
    If an entry is submitted by a client to a correct node, the node will include it in its next log
    produced.
}
\newcommand{\propLiveness}{A run of the protocol must always produce a new valid log for every correct node.}

\DeclarePairedDelimiter{\ceil}{\lceil}{\rceil}

\newcommand{\defeq}{\vcentcolon=}
\renewcommand{\emptyset}{\varnothing}
\newcommand{\union}{\cup}
\newcommand{\Union}{\bigcup}

\newcommand{\true}{\text{true}}
\newcommand{\false}{\text{false}}
\newcommand{\Bool}{\{\true,\false\}}

\newcommand{\sign}[2]{\langle#1\,|\,#2\rangle}
\newcommand{\sig}[2]{\sigma_#2(#1)}

\def\llpar{\llparenthesis}
\def\rrpar{\rrparenthesis}

\newcommand{\HOp}{\Pi}
\newcommand{\HOrd}{\mathbb{N}}
\newcommand{\HOmsg}{M}
\newcommand{\HOnullmsg}{\bot}
\newcommand{\HOmsgopt}{\HOmsg_\HOnullmsg}
\newcommand{\HOst}[1]{\mathcal{S}_{#1}}
\newcommand{\HOinit}[1]{\HOst{#1}^0}
\newcommand{\HOsend}[1]{S_{#1}}
\newcommand{\HOnext}[1]{T_{#1}}
\newcommand{\HOho}[2]{\mathit{HO}_{#1}^{#2}}
\newcommand{\HOsho}[2]{\mathit{SHO}_{#1}^{#2}}
\newcommand{\HObyz}{F}
\newcommand{\HOcorr}{C}
\newcommand{\HOsigtype}{\mathit{Sig}}
\newcommand{\HOsigs}[2]{\Sigma_{#1}^{#2}}
\newcommand{\HOextr}{\sigma}

\newcommand{\proplabel}[1]{(#1\theenumi)}
\newenvironment{enuminline}
    {\begin{enumerate*}[label=(\alph*)]}
    {\end{enumerate*}}

\makeatletter 
\let\orgdescriptionlabel\descriptionlabel
\renewcommand*{\descriptionlabel}[1]{%
  \let\orglabel\label
  \let\label\@gobble
  \phantomsection
  \edef\@currentlabel{#1}%
  \let\label\orglabel
  \orgdescriptionlabel{#1}%
}
\makeatother

\begin{document}

\title{A Formally Verified Protocol for Log Replication with Byzantine Fault Tolerance}

\author{
    Joel Wanner, Laurent Chuat, and Adrian Perrig \\
    \normalsize\textit{Network Security Group, Department of Computer Science, ETH Zurich, Switzerland}
}

\maketitle

\thispagestyle{plain}
\pagestyle{plain}

\newcommand{\locModel}[0]{312}
\newcommand{\locProof}[0]{1017}

\begin{abstract}
	\BFT protocols enable state replication in the presence of crashed, malfunctioning, or actively
malicious processes. Designing such protocols without the assistance of verification tools, however,
is remarkably error-prone. In an adversarial environment, performance and flexibility come at the
cost of complexity, making the verification of existing protocols extremely difficult. We take a
different approach and propose a formally verified consensus protocol designed for a specific use
case: secure logging. Our protocol allows each node to propose entries in a parallel subroutine, and
guarantees that correct nodes agree on the set of all proposed entries, without leader election. It
is simple yet practical, as it can accommodate the workload of a logging system such as Certificate
Transparency. We show that it is optimal in terms of both required rounds and tolerable faults.
Using Isabelle/HOL, we provide a fully machine-checked security proof based upon the Heard-Of model,
which we extend to support signatures. We also present and evaluate a prototype implementation.
\end{abstract}


\section{Introduction}

The problem of Byzantine consensus has been the subject of a considerable amount of research over
the past decades, giving rise to various \BFT ({BFT}) protocols, most notably Practical Byzantine
Fault Tolerance ({PBFT}) by~\citet{Castro2002}. In response to the publication of {PBFT}, there have
been many attempts to improve on the performance and robustness of the protocol by focusing on
different scenarios. For instance, {Zyzzyva}~\cite{Kotla2007} is designed to be especially efficient
in the absence of failures, whereas {Aardvark}~\cite{Clement2009}, on the contrary, is designed to
react gracefully when failures occur.

These protocols were designed for high-throughput, low-latency state-machine replication.
Unfortunately, this is only possible at the cost of complexity~\cite{Guerraoui2010}. The
\textsc{BFT-SMaRt}~\cite{Bessani2014} library, which implements a variant of {PBFT}, can serve as a
benchmark with almost $\num{25000}$ lines of Java code. Even in a benign fault model, where nodes
can only crash and messages may be lost but not modified, distributed systems are notoriously hard
to design and implement. In the presence of possibly malicious participants, arguing about the
correctness and security of such protocols is an even greater challenge, or in the words of
\citet{Lamport1982}: ``We know of no area in computer science or mathematics in which informal
reasoning is more likely to lead to errors than in the study of this type of algorithm.'' To
guarantee the security of such complex systems, a formal treatment is thus essential. The
traditional approach in the distributed systems community is to provide a pen-and-paper proof for
the desired properties of the protocol. At best, such proofs provide some intuition about why the
claimed properties hold, but since they lack the rigor that is required to argue about such systems,
they cannot be reasonably used as a guarantee. Past experiences, such as the Chord
protocol~\cite{Zave2012}, which had all of its hand-proved properties refuted by model checking,
have shown that proofs must be \emph{machine-checked}. Such proofs tend to be much longer and more
detailed than their hand-crafted counterparts, but manual error can be ruled out conclusively using
\emph{proof assistants}.

To the best of our knowledge, there exists no complete machine-checked proof for any
authenticated BFT protocol. Most work on verifying distributed systems has instead focused on
consensus algorithms like {Raft}~\cite{Woos2016} and {Paxos}~\cite{lamport2001paxos}, which only
tolerate benign faults. The {IronFleet} project~\cite{Hawblitzel2015} can serve as a benchmark for
the complexity of large-scale verification efforts, as it expended approximately $\num{3.7}$
person-years for the proof of a {Paxos}-based distributed system and its implementation. Due to the
tremendous complexity of a Byzantine fault model caused by the introduction of arbitrary behavior,
it is reasonable to assume that the effort of verifying a complex BFT protocol would require even
more resources.

The lack of formal verification makes general-purpose BFT protocols unsuitable for security-critical
applications, even if they have been tested and deployed in practice. To the best of our knowledge,
the only instance of complete formal verification of a BFT protocol is by \citet{Debrat2012}, who
verified two algorithms proposed by~\citet{Biely2007} using the {Isabelle/HOL}~\cite{Nipkow2002}
proof assistant. However, the properties provided by these very simple algorithms are too weak for
use in many realistic settings.

Instead of aiming to develop a general-purpose BFT system, we focus on the use case of \emph{secure
logging}, a critical component in a variety of systems: modern public-key
infrastructures~\cite{Laurie2013, Basin2014}, online voting systems~\cite{Gritzalis2002,
Chondros2015}, secure timestamping services~\cite{Massias1999, Gipp2015}, and more~\cite{Syta2015}.
In this endeavor, we make the following contributions:
\begin{itemize}
    \item
    We present \protocol, a BFT protocol designed specifically for secure log replication,
    and provide machine-checked proofs of all its properties using the {Isabelle/HOL} proof assistant.
    Our protocol model and proofs consist of approximately
    $\num[round-mode=figures,round-precision=1]{\locModel}$
    and
    $\num[round-mode=figures,round-precision=1]{\locProof}$
    lines of code, respectively, and are available online~\cite{proofs}. Although the protocol is
    simple, our verification revealed subtle flaws in its initial design, which have since been fixed.
    \item
    We extend the Heard-Of model~\cite{Charron-Bost2009} to capture the concept of digital
    signatures. Our extended model can be used to verify other BFT algorithms that make use of
    signatures.
    \item
    We evaluate the performance of a prototype implementation to demonstrate that our protocol can
    be used in practical scenarios.
\end{itemize}

\section{Background: Secure Logging}
\label{sec:background}

Logging can trivially be performed by a single server, but this server must then be trusted to (a)
accept all valid requests, (b) not remove existing entries from the log, and (c) show the same view of
the log to all clients. Verifiable data structures based on cryptographic primitives (such as Merkle
hash trees)~\cite{VerifDataStruct,dahlberg2016efficient} enable the efficient auditing of logs. This
is, most notably, the approach employed in the Certificate Transparency (CT)
framework~\cite{Laurie2013}. Verifiable logging by itself is not sufficient though, as a malicious
log server can still choose to ignore requests and show different views to different
clients~\cite{gossip2015}. A log server ignoring specific requests is particularly problematic,
because such misbehavior is hard to demonstrate and reporting it to a third party has privacy
implications~\cite{eskandarian2017certificate}.

Relying on a single server has obvious drawbacks: weakest-link security, no resilience to failure,
and no censorship resilience. On the other hand, relying on a large collection of non-synchronized
log servers makes monitoring difficult. Indeed, a client cannot simply query one CT log server to
inspect all entries related to a given domain name, for example, but must instead rely on monitors
that keep entire copies of several logs. In turn, monitors must be trusted to correctly display all
relevant entries from all trusted logs, which has revealed to be a challenge in
practice~\cite{li2019certificate}. For these reasons, we propose a protocol that allows independent
entities to maintain a single log, thus providing resilience to compromise, failure, and censorship,
while facilitating the monitoring of the log's contents by resource-limited clients.

A large majority of CT log servers accepted by Google Chrome have a ``maximum merge delay'' of 24
hours at the time of writing. This means that log servers will typically append newly submitted
certificates to their hash tree within 24 hours. In such a context, our protocol would operate on a
timescale that is perhaps unconventional for a distributed system, with each round of the protocol
lasting several hours. However, we evaluate \protocol within much smaller timeframes as well in
Section~\ref{sec:evaluation}, and find it to be able to support substantial workloads even with a
1-minute period.

\section{Problem Definition}
\label{sec:problem}

\subsection{Log Replication with Byzantine Fault Tolerance}

BFT protocols are commonly designed to achieve \emph{state-machine replication}, where processes
agree on an ordered set of incoming requests from clients, creating an input log that is equal on
all processes. Running a deterministic state machine on the log then produces the same results on
each node. The design goals in this problem space are usually low latency and high throughput,
enabling the protocol to handle a high volume of requests quickly.

This paper considers the related but slightly different problem of BFT \emph{log replication}. In
this problem, a set of $n$ nodes, of which at most $f$ may fail, periodically run a distributed
algorithm to maintain a log. There is an arbitrary number of clients in the system that can send
messages to the nodes, requesting entries to be added to the log. Moreover, the clients can obtain
the most recently created log along with an authenticator and verify the validity of its entries
locally. In order to achieve log replication, a protocol must satisfy the following properties:

\begin{LaTeXdescription}
    \item[Agreement\label{prop:main-agreement}] \propAgreement
    \item[Completeness\label{prop:main-completeness}] \propCompleteness
    \item[Liveness\label{prop:main-liveness}] \propLiveness
\end{LaTeXdescription}

This problem is different from BFT state-machine replication in three ways:

\begin{itemize}
    \item
    There exists an inherent total order on entries (e.g., alphabetical or chronological).
    Therefore, no coordination is required to determine an ordering, unlike in the state-machine
    replication problem.
    \item
    Clients are not limited to obtaining the output of a state machine. Instead, they can verify the
    integrity of the entire log, or parts of it.
    \item
    The system aggregates entries and produces new outputs in fixed intervals, not in response to
    each request.
\end{itemize}

Due to these differences, the log replication problem allows for less complex solutions, as client
requests do not need to be processed individually with low latency. Nevertheless, this problem
appears in various real-world systems, such as public-key infrastructures.

\subsection{Assumptions}

We use the standard Byzantine fault model, where up to $f$ of the $n$ nodes may crash, malfunction,
or even be actively malicious (and colluding). We call these nodes \emph{faulty}, and there are at
least $n-f$ remaining nodes that we call \emph{correct}.

The protocol relies on the following assumptions, which are common for BFT protocols:

\begin{enumerate}[label={\proplabel{A}}, leftmargin=\widthof{(A1)}+\labelsep]
    \item\label{prop:assm-correct}
    There is a correct majority of nodes, i.e., the constraint $n > 2f$ must hold.
    \item\label{prop:assm-reliable}
    Messages sent between correct nodes are neither lost nor modified, i.e., communication is
    synchronous.
    \item\label{prop:assm-keys}
    Every node has a key pair and knows the public keys of all other nodes.
\end{enumerate}

We stress that assumption~\ref{prop:assm-reliable} is attainable in practice using loose time
synchronization combined with a transport protocol that provides reliable and authenticated
communication. Moreover, if a link between two nodes fails, the nodes can simply be considered
faulty, and if the parameter $f$ is chosen large enough, the protocol will be able to continue
operating without any issues.

\section{The \protocol Protocol}
\label{sec:protocol}

\subsection{Definitions and Notation}

A log contains an ordered sequence of entries as well as an expiration timestamp. The $\fMkLog(L,
X)$ function creates a new log from the previous log $L$ and a set of new entries $X$. A log is
\emph{valid} if and only if it has not expired and is signed by $f+1$ different nodes. These
signatures serve to verify the authenticity of a log and can be computed over a digest of it (using
the $\fMkDigest(L)$ function, which may be a simple hash function, or may return the root of a hash
tree, for example). Since there are at most $f$ faulty nodes in the system, the requirement of
$f+1$ signatures ensures that they cannot collude to forge a log without obtaining a signature from
a correct node.

We use $\sig{x}{i}$ to denote a signature for $x$ created by process $i$ using its private key.
$\sign{x}{i}$ is shorthand for a signed term, i.e., $\sign{x}{i} \defeq (x, \sig{x}{i})$.
Analogously, we use $\sign{x}{W}$ for a term $x$ that is signed by a set of nodes $W$, which are
called witnesses.

\subsection{Protocol Overview}

The protocol is organized into rounds of communication, which can be implemented in asynchronous
settings assuming loose time synchronization.

A key insight into how the protocol can be kept simple while still achieving strong security
properties is that it does not require leader election. This concept allows low-latency BFT
protocols to process requests quickly by designating one process as a leader, also called
\emph{primary}. While this can reduce the number of messages required, electing a leader also
introduces a high amount of complexity, as the system needs to be able to handle cases where the
current primary crashes or is actively malicious. For this purpose, processes can initiate a ``view
change'' phase to convince others to hold a new election.

At the core of \protocol lies the \algSub subroutine, a distributed consensus algorithm. While this
subroutine is also based on a primary, we avoid the process of leader election by running $n$
instances in parallel, such that each node is the leader of exactly one thread. The goal of this
technique is to achieve \emph{interactive consistency}, which \citet{Pease1980} define as follows.
Each node $i$ chooses an initial value $v_i$, and the following properties must hold:

\begin{enumerate}[label={\proplabel{IC}}, leftmargin=\widthof{(IC1)}+\labelsep]
    \item\label{prop:ic-agreement}
    All correct nodes agree on the same vector $V$ of $n$ values.
    \item\label{prop:ic-validity}
    For a correct node $i$, all correct nodes agree on $i$'s initial value: $V_i = v_i$.
\end{enumerate}

\begin{figure}[tb]
    \centering
    \begin{tikzpicture}

\def\roundsize{0.4cm}
\def\logheight{0.4cm}
\def\runsize{9*\roundsize}
\def\padding{0.2cm}

\tikzset{
    round/.style={
        rectangle,
        draw,
        minimum width=\roundsize,
        minimum height=\roundsize
    },
    lround/.style={round, minimum width=\runsize-5*\roundsize},
    faded/.style={draw=gray},
    log/.style={
        rectangle,
        draw=dark-gray,
        fill=light-gray,
        minimum height=\logheight,
        minimum width=\runsize+\roundsize
    }
}

\foreach \i in {1,...,2}{
    \foreach \j in {1,...,5}{
        \ifnum \i=1
            \def\cls{faded}
        \else
            \def\cls{}
        \fi
        \node[round,\cls] at (\i*\runsize+\j*\roundsize,0) (R\i\j) {};
    }
}

\node [lround, right] at (R15.east) {\scriptsize\PhaseA};

\node at (R21) {$1$};
\node at (R22) {$2$};
\node at (R23) {$3$};
\node at (R24) {\scriptsize\ldots};

\node [left] at (R11.west) {\ldots};
\node [lround, faded, anchor=west] at (R25.east) (R26) {};
\node [right] at (R26.east) {\ldots};

\draw [line width=1pt] (R21.south west) -- (R21.north west);
\draw [line width=1pt] (R24.south east) -- (R24.north east);
\draw [line width=1pt] (R25.south east) -- (R25.north east);

\draw [hbrace-tiny] (R24.south east) -- (R21.south west)
    node [midway, below] (B){\scriptsize\PhaseB};
\node [below=-4pt] at (B.south) {\scriptsize{($f+1$ rounds)}};
\draw [hbrace-tiny] (R25.south east) -- (R25.south west)
    node [midway, below, xshift=6pt] {\scriptsize\PhaseC};

\begin{scope}
    \clip (\runsize,0) rectangle (3*\runsize+\roundsize,2*\logheight+\roundsize+3*\padding);
    \node [log, anchor=south east] at ($(R15.north east)+(\roundsize,\padding)$)
        (L1) {\scriptsize $L_{i-2}$};
    \node [log, anchor=south west] at ($(R15.north east)+(0,2*\padding+\logheight)$)
        (L2) {\scriptsize $L_{i-1}$};
    \node [log, anchor=south west] at ($(R25.north east)+(0,\padding)$)
        (L3) {\scriptsize $L_i$};
    \draw [thick, dotted, draw=gray] (R25.north east) -- ++(0,2*\logheight+3*\padding);
\end{scope}

\end{tikzpicture}
    \caption{
        An overview of the protocol operation. Each white rectangle represents a round of
        communication. Logs are characterized by their validity period. The exact definition of a
        validity period is left to application developers. Round lengths can also be tuned to each
        application.
    }
    \label{fig:phases}
\end{figure}
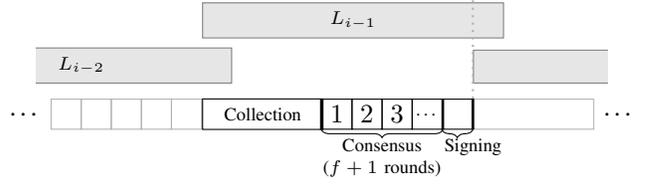

\begin{algorithm}[t]
    \caption{Main protocol}
    \label{alg:main}
    \begin{algorithmic}[1]
	    \Procedure{\algMain}{}
	        \Statex \emph{Code for node $i$}
	        \Statex \emph{\PhaseA{} phase:}
	        \State $X \gets$ collect entries from clients
	        \Statex \emph{\PhaseB{} phase:}
	        \State $\{X_1,\ldots,X_n\} \gets \{\algSub(p, X): 1 \leq p \leq n\}$
	        \State $X \gets X_1 \union \ldots \union X_n$
	            \label{line:sub-ic}
	        \Statex \emph{\PhaseC{} phase:}
	        \State $L \gets$ log from previous epoch
	        \State $L' \gets \fMkLog(L, X)$
	        \State $h \gets \fMkDigest(L')$
	        \State broadcast $\sig{h}{i}$
	        \State $\Sigma \gets$ receive $\theta-1$ valid signatures for $h$
	        \State publish $(L', \Sigma \union \{\sig{h}{i}\})$
	    \EndProcedure
	\end{algorithmic}
\end{algorithm}

Our protocol operation is specified in detail in Algorithm~\ref{alg:main} and illustrated in
Figure~\ref{fig:phases}. It consists of three phases:

\begin{enumerate}
    \item
    \textbf{\PhaseA} ($1$ round)\\
    Clients can send requests for entries to be added to the log, where each request should be sent
    to $f+1$ distinct nodes to  guarantee that at least one correct node receives it. Each node
    stores all entries that it has received and that have not been added locally. During this phase,
    no communication takes place between the nodes.

    \item
    \textbf{\PhaseB} ($f+1$ rounds)\\
    With the set $X$ of all cached entries from the previous phase as input, the \algSub subroutine
    is run in parallel $n$ times. In each of these executions, a different node acts as the primary.
    The \algSub subroutine is explained in more detail in Section~\ref{sec:protocol-sub}, but for
    the sake of this overview, we can assume that it achieves interactive consistency for initial
    values $X$ and the result vector is $(X_1,\ldots,X_n)$. This implies that after
    line~\ref{line:sub-ic}, all correct nodes obtain the same value for $X$, and their collected
    entries from the previous phase are contained in $X$.

    \item
    \textbf{\PhaseC} ($1$ round)\\
    Using the log from the previous run and the union of all new entries, each node constructs a new
    log.

    After the new log $L'$ is constructed, each node broadcasts a signature for $\fMkDigest(L')$ to
    all other nodes.%
    \footnote{
    The log needs not be sent along with the signature, as the other correct nodes will have
    constructed the same log $L'$.
    }
    Each node then collects signatures from $f$ other participants that have also constructed the
    same log and publishes it along with all signatures.
\end{enumerate}

In the following subsection, we describe in detail the \algSub subroutine, which is the main
building block of the protocol, and explain how the \phaseB phase achieves interactive consistency.

\subsection{\PhaseB Phase}
\label{sec:protocol-sub}

\begin{algorithm}[t]
    \caption{\PhaseB phase}
    \label{alg:sub}
    \begin{algorithmic}[1]
	\Procedure{\algSub}{$p,X$}
	\Statex \emph{Code for node $i$:}
	\If {$i = p$} \Comment{node is primary}
	    \State broadcast $\sign{X,p}{i}$
	    \State return $X$
	\Else \Comment{node is responder}
	    \State $P \gets \emptyset$ \Comment{witnessed values}
	    \State $d \gets \emptyset$ \Comment{decision value}
	
	    \For {rounds $r = 1,\ldots,f+1$}
	        \State $M \gets$ receive messages
	        \State $M' \gets \{\sign{x,p}{W} \in M.~ p \in W \land |W| \ge r\}$
	            \label{line:sub-extract}
	            \State $P' \gets \{x.~ \exists W.~ \sign{x,p}{W} \in M'\}$
	        \If {$P' \setminus P \neq \emptyset$}
	            \If {$|P \union P'| = 1$} \label{line:sub-pcheck}
	                \State $d \gets$ the only element of $P'$
	            \Else
	                \State $d \gets \emptyset$
	            \EndIf
	            \State $S \gets \emptyset$ \label{line:sub-forward_begin}
	            \For {$x \in P' \setminus P$}
	                \State $\sign{x,p}{W} \gets$ any element of $M'$
	                \State $S \gets S \union \{\sign{x,p}{W \union \{i\}}\}$
	                    \label{line:sub-forward_end}
	            \EndFor
	            \State multicast $S$ to other responders
	        \EndIf
	        \State $P \gets P'$
	    \EndFor
	    \State \Return $d$
	\EndIf
	\EndProcedure
	\end{algorithmic}
\end{algorithm}

For the sake of abstraction, we will refer to a set of entries as a \emph{value} and note that
\algSub is not limited to a specific type of value, but can be used more generally. The algorithm
must satisfy the following two properties:

\begin{enumerate}[label={\proplabel{\prefixSubProps}}, leftmargin=\widthof{(C1)}+\labelsep]
    \item\label{prop:sub-agreement}
    All correct nodes return the same value.
    \item\label{prop:sub-validity}
    If the primary $p$ is correct, all correct nodes return the primary's input value $X$.
\end{enumerate}

If these properties hold for all parallel executions of \algSub, it follows that the \phaseB phase
satisfies \ref{prop:ic-agreement} and \ref{prop:ic-validity}. The main challenge in designing the
algorithm does not lie in securing executions in which the primary is correct, but in preventing a
faulty primary from causing disagreement between correct nodes. This is especially difficult because
the primary may be actively malicious and colluding with all other faulty nodes.

Informally, the algorithm works as follows. Only the primary can propose new values, which it does
by signing them. When a process receives such a value, it forwards the value to other nodes and
testifies to witnessing it by appending its own signature. This is necessary for two reasons:
\begin{enuminline}
    \item to inform nodes about the primary's value in case the message in the initial round was
        lost, and
    \item to detect \emph{equivocation} by the primary (i.e., sending different values in an attempt
        to create different views).
\end{enuminline}

However, this technique alone is not enough to prevent attacks: A malicious primary could collude
with responders and attempt to introduce a new value to some nodes in the last round, so the
receivers will not have time to share it. To prevent this, correct nodes also apply the following
policy: In round $r$, received values that have fewer than $r$ witnesses are discarded. Any values
sent in the last round must therefore be signed by $f+1$ nodes in order to be accepted, i.e., at
least one correct node must have witnessed it previously.

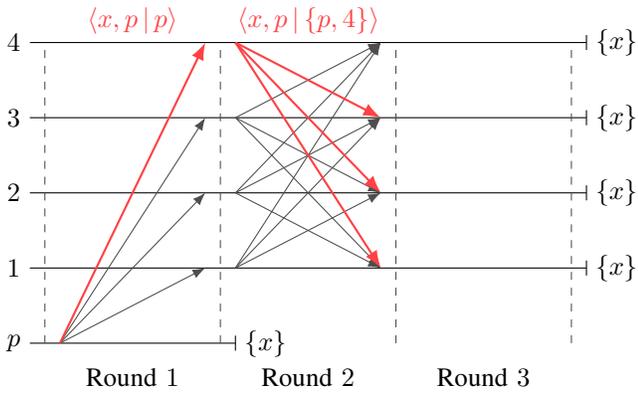
\begin{figure}
    \centering
    \begin{tikzpicture}

\tikzset{
    arrow/.style={-Latex},
    faded/.style={draw=darker-gray},
    highlight/.style={draw=color1, thick},
    capped/.style={
        postaction={decorate,
            decoration={markings,
            mark=at position 1 
                with {
                    \draw (0,0.1) -- (0,-0.1);
                }
            }
        }
    },
}

\def\n{4}
\def\width{7}
\def\height{4}
\pgfmathsetmacro\row{\height/(\n)}
\def\margin{0.2}
\def\rounds{3}
\def\step{\width/\rounds}

\foreach \i in {0,...,\n}{
    \ifnum \i=0
        \def\l{p}
        \def\d{\step} 
    \else
        \def\l{\i}
        \def\d{\width}
    \fi
    \node[left] at (-\margin,\i*\row) (\i) {$\l$};
    \draw[capped] (\i) -- (\d+\margin,\i*\row) node[right] {$\{x\}$};
}

\foreach \r[evaluate=\r as \evalr using int(\r+1)] in {0,...,\rounds}{
    \draw[dashed, faded] (\r*\step,0) -- ++(0,\height);
    \ifnum \r<\rounds
        \node[below] at (\r*\step+\step/2,-0.2) {Round $\evalr$};
    \fi
}

\foreach \i in {1,...,\n}{
    \ifnum \i=\n \def\cls{highlight} \else \def\cls{faded} \fi
    \draw[arrow,\cls] (\margin,0) to (\step-\margin,\i*\row);
    \foreach \j in {1,...,\n}{
        \ifnum \j=\i \else
            \draw[arrow,\cls] (\step+\margin,\i*\row) to (2*\step-\margin,\j*\row);
        \fi
    }
}

\node[above,color1] at (0.5*\step,\height) {$\sign{x,p}{p}$};
\node[above,color1] at (1.5*\step,\height) {$\sign{x,p}{\{p,4\}}$};

\end{tikzpicture}
    \caption{
        The normal case of an execution of \algSub with a correct primary $p$. There are $n=5$
        nodes, of which $f=2$ may be faulty, which requires $f+1=3$ rounds of communication. In the
        first round, $p$ broadcasts its initial value, and in the second round, each responder
        decides on this value and forwards it to others. No further communication is required in the
        last round.
    }
    \label{fig:normal-case}
\end{figure}

After introducing the main concepts used in \algSub, we now explain the subroutine in more detail
and refer to the specification in Algorithm~\ref{alg:sub}. In most situations, the execution of
\algSub is simple and requires only two rounds of communication. Such a normal case is depicted in
Figure~\ref{fig:normal-case}.

\subsubsection{Primary}

The procedure for the primary $p$ is simple: $p$ broadcasts its initial value in the first round and
returns. A term of the form $\sign{x,p}{i}$ represents the assertion ``node $i$ testifies to having
witnessed value $x$ from primary $p$''. It is important to include $p$ in the signature, as this
prevents replay attacks using signatures received in different threads.

\subsubsection{Responders}

The other nodes, which we call responders, keep in their state the set $P$ containing all values
witnessed from the primary, as well as their current decision value $d$. In the following, we
describe the behavior of a responder node with identifier $i$. During each round $r$, node $i$
listens for any messages of the form $\sign{x,p}{W}$ and stores them in the set $M$ (discarding any
invalid signatures). After a timeout, $i$ advances to line~\ref{line:sub-extract} and selects from
$M$ all messages that are
\begin{enuminline}
    \item signed by the primary, and
    \item witnessed by a total of $r$ nodes (including the primary).
\end{enuminline}
From these messages $M'$, all distinct values are extracted and stored in $P'$, which now contains
all values that have been accepted this round.

If $i$ has accepted any new values this round (i.e., $P' \setminus P \neq \emptyset$), it adjusts
its decision based on the values it has accepted so far (line~\ref{line:sub-pcheck}). If $|P \union
P'| = 1$ holds, $i$ stores the new value in $P'$ as its current decision. If there are more values
in the two sets, $i$ has proof of misbehavior by $p$ because a correct primary will never send more
than one value. In this case, $i$ decides on the default value, which is the empty set.

In both of these two cases, $i$ needs to testify to having witnessed the new values in $P' \setminus
P$. For this purpose, a message $m$ is assembled containing all new values with the corresponding
signatures (lines~\ref{line:sub-forward_begin}--\ref{line:sub-forward_end}). For each new value $x$,
$i$ adds a signature of its own to the message. $m$ is then broadcast to all other responders.

Finally, after all $f+1$ rounds have completed, $i$ concludes the \phaseB phase by returning its
current decision $d$.

\section{Analysis}
\label{sec:analysis}

\subsection{Informal Security Analysis}
\label{sec:analysis-security}

The goal of the following informal arguments is to provide intuition as to why our protocol
satisfies the claimed properties. A more rigorous treatment using formal proofs is presented in
Section~\ref{sec:verification}.

The security arguments are structured such that the properties of \algSub are discussed first. By
building on these, we argue that \algMain achieves agreement, completeness, and liveness. This is
also the structure that our formal proof follows, as it allows for a more modular argument by
treating the two algorithms separately, with the properties~\ref{prop:sub-agreement}
and~\ref{prop:sub-validity} as the only interface between them.

\subsubsection{\PhaseB phase}

To show the validity property~\ref{prop:sub-validity}, we can assume that the primary $p$ is
correct. In this case, all the primary does is broadcast its value $X$ to all responders and decide
on it. Since channels between correct nodes are reliable by Assumption~\ref{prop:assm-reliable}, all
correct responders will receive $X$. Any other values will not be accepted by correct nodes, since
the algorithm discards all values that are not signed by the primary. Therefore, all correct nodes
will decide on $X$ and validity is satisfied. This also implies the agreement property
\ref{prop:sub-agreement} for the case of a correct primary.

We now show that agreement is also guaranteed in the presence of a faulty primary, which follows
from two observations:
\begin{enuminline}
    \item When a correct node accepts a new value, all other correct nodes will have accepted it in
    the next round. This is implied by a combination of~\ref{prop:assm-reliable} with the fact that
    nodes sign and share all new values they accept.
    \item No new values can be introduced into the system in the last round. This follows from the
    property that in this round, values require $f+1$ signatures to be accepted, so any new value
    must have been signed (and therefore shared with all other responders) by a correct node in an
    earlier round.
\end{enuminline}

\subsubsection{\protocol}

Property~\ref{prop:sub-agreement} ensures that all correct nodes obtain the same set $X$. Along with
the fact that $\fMkLog(L, X)$ is a deterministic function, i.e., imposes some known order on the
entries in $X$, this implies that all correct nodes produce the same log $L'$. There are at least
$n-f \ge f+1$ correct nodes by assumption \ref{prop:assm-correct}, and because
\ref{prop:assm-reliable} implies reliable channels between them, each of the correct nodes obtains
signatures for $L'$ from at least $f+1$ nodes (including themselves). This satisfies the
\ref{prop:main-liveness} property.

In order to satisfy the \ref{prop:main-agreement} property, we not only need to show that all
correct nodes produce the same log, but also that there can exist no other log. This follows from
the assumption that there are at most $f$ faulty nodes, and since $f+1$ are required for a valid
log, this implies that they need the signature of at least one correct node to forge a log. However,
since all correct nodes produce the same log, the faulty processes will not be able to create a
different log.

Finally, we give some intuition about~\ref{prop:main-completeness}. Let $i$ be a correct node to
which the client sent its request. Then, $i$ will include the requested entry $x$ in its proposal
$X$, which is given as input to the \algSub subroutine. Using~\ref{prop:sub-validity}, we have that
$x$ is contained in the set $X_i$. When combined with the agreement properties above, this implies
that all correct nodes will include $x$ in their new log.

\subsubsection{Denial of Service}
\label{sec:dos}

A common problem with BFT algorithms is that malicious actors may attempt to introduce a large
number of values into the system in order to decrease the effective throughput of legitimate
requests. Concretely, an attacker may send an excessive number of requests to a node such that it is
not able to broadcast their value to all receivers within the duration of a round.

Mitigation against such attacks is specific to the application deployed on top of the protocol and
is thus considered out of scope, but we envision countermeasures like rate limiting, client request
authentication and duplicate suppression. These are all common techniques to strengthen resilience
of systems to denial-of-service attacks.

\subsection{Relation to Theoretical Bounds}

\citet{Pease1980} showed that Byzantine consensus for at most $f$ faults requires at least $f+1$
rounds of communication. An intuition for this bound is as follows. A faulty node may send different
values to different nodes (i.e., perform a split-world attack, also called equivocation). This is
not necessarily a problem, since correct nodes can communicate with each other in the next round to
try to reach the same state. However, we want to avoid that faulty nodes keep proposing new values
ad infinitum. Therefore, we must require that messages be signed by more nodes in further rounds
(i.e., at least $i$ nodes in round $i$). Even with the above requirement, in any round $i \le f$,
faulty nodes can still make one correct node accept a new value. That correct node must be able to
show the new value to other correct nodes in the next round, and the only thing it can do is add its
own signature. Therefore, in round $i+1$, nodes should accept messages with just $i+1$ signatures.
As a result, in the worst case, consensus will only be reached in round $f+1$.

Our protocol attains this lower bound because consensus is reached entirely within the \phaseB
phase, which consists of $f+1$ rounds. The last round (the \phaseC phase) only serves to collect
enough signatures to publish the list that the nodes have agreed upon.

Moreover, \citet{Pease1980} proved that if signatures can be used, consensus is possible for any $n
> f$. While we assume that $n > 2f$, the \phaseB phase also reaches consensus for the weaker bound
of $n > f$ (which can be shown by modifying the assumptions in our {Isabelle/HOL} proof). However,
it is intuitively clear that the secure log replication problem is not solvable in the case of $n
\le 2f$: Consensus can be reached among the servers, but clients cannot reliably obtain the correct
result in the presence of adversaries. Since a client cannot distinguish between faulty and correct
nodes, the only way to ensure that the faulty participants do not forge a result from the protocol
is to require a quorum of nodes to verify it. This is the purpose of requiring $\theta = f+1$
signatures for a log to be valid. Therefore, the system always requires a correct majority of nodes,
and the $n > 2f$ bound holds.

We thus conclude that our protocol is optimal with respect to both the number of rounds and the
number of faulty nodes tolerated, considering the theoretical bounds on the problem of secure log
replication.

\section{Evaluation}
\label{sec:evaluation}

We have implemented a prototype of \protocol in Go consisting of approximately $\num{600}$ lines of
code and used it in a series of experiments to evaluate the practical performance of the protocol.
The experiments were conducted in a setting similar to that of Certificate Transparency (CT). In our
setup, a varying number of nodes coordinate to keep an identical log of certificate entries. The
workload is simulated using randomly generated strings of length equal to the average size of a CT
log entry, which we determined to be approximately $\num{1570}$ bytes. The period length (duration
of a protocol run) was chosen as one minute. In order to determine if \protocol can support a
realistic workload, we compare our results to the average throughput of one of the largest CT logs
to date, Google Argon 2019~\cite{ct-stats}.

The nodes were run in a virtual network created using Mininet on a single machine with two $8$-core
Intel Xeon E5-2680 processors running at $\SI{2.70}{GHz}$ and with $\SI{32}{GB}$ of RAM. In the
virtual topology, each node was connected to the other hosts by a link with capacity
$\SI{100}{Mb/s}$.

\subsection{Scaling}

\begin{figure}
    \centering
    \begin{tikzpicture}
    	\footnotesize
        \begin{axis}[
            title={$f=2$},
            xlabel={Number of nodes $n$},
            ylabel={Average throughput $\rho$ $[\si{KB/s}]$},
            xmin=4.5, xmax=25.5,
            ymin=0, ymax=750, ytick distance=100,
            legend pos=north east,
            width=\columnwidth, height=5cm
        ]
        \addplot plot[error bars/.cd, y dir=both, y explicit]
            table[x=n, y=y, y error=e, col sep=comma]
            {exp1.csv};
        \addplot[mark=none, samples=2, domain=0:50] {40.55};
        \legend{\protocol prototype, Google Argon 2019 log}
        \end{axis}
    \end{tikzpicture}
    \caption{
        Average throughput of our Logres prototype with $f = 2$, depending on the number of replicas
        $n$. For comparison, we indicate the average throughput of the (non-replicated) Google Argon
        2019 CT log server, which processes $\num{92824}$ certificates per hour on average at the
        time of writing.
    }
    \label{plt:exp1}
\end{figure}
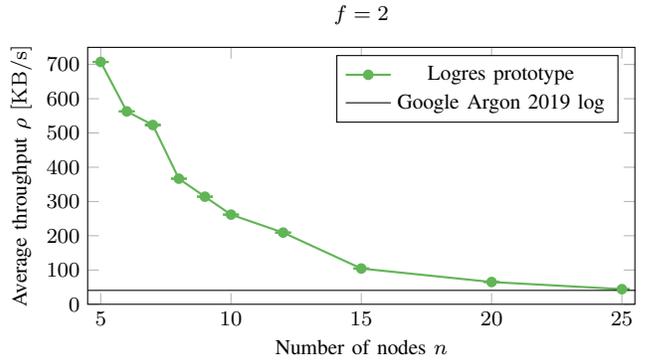

In this experiment, we measure the throughput of the system as the number of nodes increases, in
order to evaluate the scalability of \protocol. The results are displayed in Figure~\ref{plt:exp1}.
We observe that performance diminishes as more nodes are added, which can be explained by the fact
that messages need to be broadcast to other nodes. The uplink capacity limits the amount of data
that can be sent during a round, and if this capacity must be distributed among more nodes, the
maximum possible throughput decreases. Despite these effects, we observe that our system is able to
support the load of a real-world CT log with up to $25$ nodes.

We used a constant number $f = 2$ of tolerated faulty nodes in this experiment. This parameter
constitutes a typical trade-off between performance and security (as discussed in
Section~\ref{sec:exp-f}).

\subsection{Malicious Nodes}
\label{sec:exp-f}

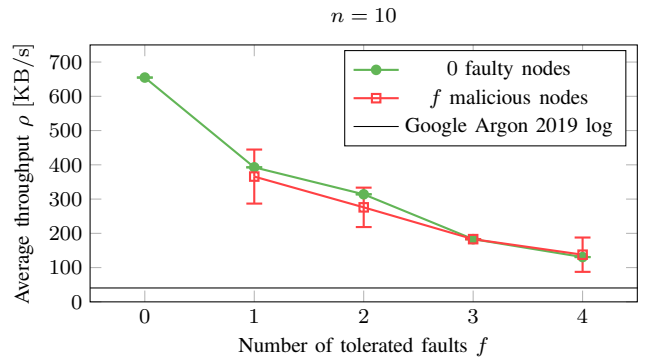
\begin{figure}
    \centering
    \begin{tikzpicture}
    	\footnotesize
        \begin{axis}[
            title={$n=10$},
            xlabel={Number of tolerated faults $f$},
            ylabel={Average throughput $\rho$ $[\si{KB/s}]$},
            xmin=-0.5, xmax=4.5,
            ymin=0, ymax=750,
            xtick=data, ytick distance=100,
            width=\columnwidth, height=5cm
        ]
        \addplot plot[error bars/.cd, y dir=both, y explicit]
            table[x=n, y=y, y error=e, col sep=comma]
            {exp2a.csv};
        \addplot plot[error bars/.cd, y dir=both, y explicit]
            table[x=n, y=y, y error=e, col sep=comma]
            {exp2b.csv};
        \addplot[mark=none, samples=2, domain=-1:5] {40.55};
        \legend{$0$ faulty nodes, $f$ malicious nodes, Google Argon 2019 log}
        \end{axis}
    \end{tikzpicture}
    \caption{
        The impact of increasing the parameter $f$ for a fixed $n$ in the case where all nodes are
        correct, compared to a simulation of malicious behavior.
    }
    \label{plt:exp2}
\end{figure}

The parameter $f$ determines how many faults can be tolerated and may be chosen up to a maximum of
$\ceil{\frac{n}{2}}-1$. However, larger values of $f$ are also detrimental to performance. Moreover,
even though we have proven the protocol secure for these choices of $f$, adversaries may still be
able to deteriorate the performance of the system by causing communication overhead. The purpose of
this experiment is to measure the impact of these factors. For this purpose, we use a setup with $n
= 10$ nodes, where measurements are conducted both with correct nodes and with a simulation of
malicious behavior. The results are displayed in Figure~\ref{plt:exp2}.

\subsubsection{Correct nodes}

In the absence of faulty nodes, the parameter $f$ has two effects on performance: The \phaseB phase
consists of $f+1$ rounds, so the rounds become shorter as $f$ is increased. This limits the total
size of entries that can be proposed with the same network bandwidth. Secondly, the clients are also
affected by this parameter. In order to ensure that a request is not dropped by faulty nodes, a
client should replicate it to $f+1$ nodes, which amplifies the total data volume that needs to be
exchanged during the protocol run.

\subsubsection{Malicious behavior}

The simulation of malicious behavior is based on a number of observations. The first is that in
executions of \algSub with a correct primary, an adversary cannot cause the correct nodes to deviate
from their regular behavior. This is because any value that is not signed by the primary is rejected
immediately. Moreover, a malicious primary can be blocked immediately by the responders once
equivocation is detected, since the signatures created can be used to hold the primary accountable.
Finally, we have discussed in Section~\ref{sec:dos} how denial-of-service attacks should be
mitigated by the application running on top of \protocol.

From these observations and the properties proved for the protocol, we can rule out many behavior
patterns that allow the correct nodes to block the adversaries. The attack strategy we implemented
works as follows: The faulty nodes generate some values, share them among each other and sign them.
These values are then sent to the correct nodes in a later round. Since the messages carry more
signatures than usual, this causes some communication overhead at the correct nodes.

In our implementation, signatures are $\num{512}$ bytes long and thus small in comparison to the
large volume of values that are exchanged. Therefore, we do not expect this attack to have a great
impact. Our hypothesis is confirmed in Figure~\ref{plt:exp2}, which shows no significant difference
in average system throughput between the normal case and the attack scenario. We therefore conclude
that \protocol performs well even in the presence of malicious participants that actively try to
sabotage the protocol execution.

\subsection{Latency}

\begin{table}
    \centering
    \caption{Time required for a request to be processed, depending on the number of tolerated
    faults $f$, with $n=5$ nodes.}
    \label{plt:exp3}
    \begin{tabular}{lll}
        \toprule
        $f$ & Lower Bound & Measured Latency \\
        \midrule
        0 & \SI{40.34}{ms} & \SI{44}{ms}\\
        1 & \SI{60.51}{ms} & \SI{66}{ms}\\
        2 & \SI{80.68}{ms} & \SI{88}{ms}\\
        \bottomrule
    \end{tabular}
\end{table}

Finally, we investigate the minimum latency attainable by the protocol. For the previous
experiments, the system was configured with a period length of one minute, but a shorter interval
can be chosen for applications that are subject to stricter latency requirements.

We will determine the minimum time that is required for a request to be processed, i.e., from the
moment the request is received by a server until it is added to a new version of the log. A lower
bound on the round length can be computed using the network parameters between nodes and the size of
requests. In this experiment, we choose links with a latency of $\SI{20}{ms}$ and a bandwidth
$\SI{100}{Mb/s}$. Again, we use $\SI{1570}{B}$ values, which results in a minimum bound on round
length of approximately $\SI{20.17}{ms}$ (the time required for transmission of a single value).

To measure the actual latency of the system, the clients submit a single request for each period,
and we decrease the round length successively until the minimum value (for which the requested value
is still processed reliably) is reached. We expect the measurements to exhibit slightly larger
values than the theoretical lower bound due to the computational overhead due to message processing
and signature verification.

The results from the experiment are displayed in Table~\ref{plt:exp3} for $5$ nodes and different
values of $f$. The results show that our implementation is able to achieve small latencies, close to
the theoretical lower bound.

\section{Verification}
\label{sec:verification}

Our verification is based on the Heard-Of model, which was introduced by \citet{Charron-Bost2009}.
We first provide a brief overview of this model and how we extended it to account for signatures.
Finally, this section contains a brief overview of our results and experiences with constructing the
proof. We refer to the appendix for more details about the protocol formalization
(Appendix~\ref{sec:verification-protocol}) and the proof methodology
(Appendix~\ref{sec:verification-isabelle}) we used in the process.

\subsection{The Heard-Of Model}

The Heard-Of model provides a framework for modeling consensus algorithms in a lock-step model of
the system, where communication is divided into rounds. As the original model only accounted for
benign faults, \citet{Biely2007} extended it to capture malicious behavior. Our proofs are based on
an implementation of this model in Isabelle/HOL by \citet{Debrat2012}.

In this context, we will refer to nodes more generally as processes. The set of $n$ processes is
represented by $\HOp$, and $\HOmsg$ denotes the set of possible messages. We define $\HOmsgopt
\defeq \HOmsg \union \{\HOnullmsg\}$, where $\HOnullmsg$ stands for the absence of a message. Each
$i \in \HOp$ is associated with a process specification $P_i =
(\HOst{i},\HOinit{i},\HOsend{i},\HOnext{i})$ consisting of the following components:
\begin{itemize}
    \item $\HOst{i}$:
        the set of $i$'s \emph{states}, and $\HOinit{i}$: the set of possible \emph{initial states}
        of process $i$.

    \item $\HOsend{i}: \HOrd \times \HOst{i} \times \HOp \rightarrow \HOmsg$:
        the \emph{message sending function}, where $\HOsend{i}(r,s,j)$ denotes the message sent by
        $i$ to $j$ in round $r$, given that $i$ is in state $s$.

    \item $\HOnext{i}: \HOrd \times \HOst{i} \times \HOmsgopt^{\HOp} \times \HOst{i} \rightarrow
        \Bool$: the \emph{next-state predicate}, where $\HOnext{i}(r,s,\mu,s')$ evaluates to $\true$
        if and only if $s'$ is a state that $i$ can reach from state $s$ in round $r$, given the
        vector $\mu$ of messages that were received by $i$ in round $r$.
\end{itemize}
The collection of process specifications $P_i$ is called an \emph{algorithm} on $\HOp$.

A \emph{run} of such an algorithm is defined, for each process $i$, by a sequence of states
$s_i^0,s_i^1,\ldots$ that satisfies the following conditions:
\begin{itemize}
    \item
    $s_i^0$ is a valid initial state: $s_i^0 \in \HOinit{i}$
    \item
    For each $r \in \HOrd$, it holds that the next-state predicate
    $\HOnext{i}(r,s_i^r,\mu_i^r,s_i^{r+1})$ is true for a message vector $\mu_i^r$ collected from
    the messages sent by the other processes.
\end{itemize}
We now define $\mu_i^r$, which not only depends on the specification of the algorithm, but also on
the communication assumptions. For each process $i$ and each round $r$, we define two subsets of
$\HOp$:
\begin{itemize}
    \item $\HOho{i}{r}$:
    the \emph{heard-of set}, which places the following constraint on the message vector:
    \begin{equation*}
        j \in \HOho{i}{r} \iff \mu_i^r[j] \neq \HOnullmsg
    \end{equation*}
    This condition states that some message arrives at $i$ from each process in $\HOho{i}{r}$, which
    may or may not conform to the process specification.
    \item $\HOsho{i}{r}$:
    the \emph{safe heard-of set}, the set of processes whose message arrives unchanged:
    \begin{equation*}
        j \in \HOsho{i}{r} \iff \mu_i^r[j] = \HOsend{j}(r,s_j^r,i)
    \end{equation*}
    The message received by $i$ is therefore guaranteed to be the one given by the protocol
    specification for $j$.
\end{itemize}

\subsection{Extension to Signatures}
\label{sec:verification-sho_model}

The standard Byzantine fault model can be encoded by defining a subset $\HObyz \subseteq \HOp$ of
Byzantine faulty nodes with $|\HObyz| \leq f$ and choosing the heard-of sets as
$\HOsho{i}{r} = \HOp \setminus \HObyz \subseteq~\HOho{i}{r}$. This corresponds to the assumption
that the messages of correct nodes arrive according to process specification, and that faulty nodes
may send any message in $\HOmsg$ or none at all. However, this property is generally too weak for a
protocol that relies on digital signatures. This cryptographic primitive is commonly used under the
assumption that signatures are unforgeable, which implies that faulty nodes must not be allowed to
simply send arbitrary elements from the set $\HOmsg$.

Instead, we introduce a restriction on messages sent by faulty nodes that encodes the unforgeable
nature of signatures. While this could also be achieved by modeling Dolev-Yao intruder knowledge,
we chose a simpler approach that requires fewer modifications to the original model. Let
$\HOsigtype$ denote the set of signatures and for each process $i$, let $\HOextr_i: \HOp \mapsto
2^{\HOsigtype}$ be a function that given a message $m$, extracts the set of signatures contained in
$m$ that are signed by $i$.  Furthermore, $\HOsigs{i}{r}$ is the set of all signatures sent by $i$
in round $r$:
\begin{equation*}
    \HOsigs{i}{r} \defeq \Union_{j \in \HOp} \HOextr_i(\HOsend{i}(r,s_i^r,j))
\end{equation*}
From this, we define $M_r$ as the set of all messages that can be constructed from signatures sent
out until round $r$:
\begin{equation*}
    \HOmsg_r \defeq \{m \in M:~ \forall i \in \HOcorr_r.~ \HOextr_i(m) \subseteq \Union_{r'=0}^{r}
    \HOsigs{i}{r}\}
\end{equation*}
where $\HOcorr_r$ denotes the set of processes that have not failed yet, i.e., processes that have
been contained in all secure heard-of sets up to this round. Intuitively, this definition means that
an agent that has stored all messages up to round $r$ can use its knowledge to construct any message
in $M_r$ (e.g., including signatures received from correct nodes). Note that malicious nodes are not
in the set $\HOcorr_r$, and therefore signatures from them may appear in any messages. This allows
for collusion between attackers, which may share their private keys over out-of-band channels.

Using these definitions, we can finally define the restriction on message vectors $\mu_i^r$ for all
rounds $r$ and processes $i$ and $j$:
\begin{equation*}
    \mu_i^r[j] \in \HOmsg_r \union \{\HOnullmsg\}
\end{equation*}
In summary, our extension of the HO model encodes the standard assumption that signatures are
unforgeable, while still allowing faulty processes to sign arbitrary values and colluding with each
other.

\subsection{Verification Results}

Our Isabelle/HOL verification of all security properties claimed in this paper is available
online~\cite{proofs}. It consists of approximately
$\num[round-mode=figures,round-precision=1]{\locModel}$ lines of code for the protocol model and
$\num[round-mode=figures,round-precision=1]{\locProof}$ lines of code for the proofs.

The verification process revealed two flaws in previous versions of the protocol. The first was
based on missing isolation between instances of the \algSub subroutine, which are invoked in
parallel by \protocol. Since signatures were not bound to a specific subroutine, a sophisticated
attacker could have been able to replay legitimate signatures obtained from one thread in another
thread, thereby preventing nodes from creating a valid log. This vulnerability was fixed by binding
all signatures from a \algSub routine to the corresponding primary.

In the second attack, an adversary could cause some nodes to exit from the \phaseB phase prematurely
and thus again prevent the creation of a valid log.

These insights from our proofs provide further anecdotal evidence towards the conclusion that proofs
must be machine-checked to provide strong guarantees. The vulnerabilities described above were not
discovered during pen-and-paper construction of the proof, but only during the process of
mechanizing it. This was detected when the proof assistant rejected some proof steps that were
believed to be correct after informal reasoning.

The lessons learnt in this process and the practices we are recommending in
Appendix~\ref{sec:verification-isabelle} have also been identified in other verification projects
\cite{Klein2009, Woos2016}. Since the advent of large-scale mechanized proofs, significant progress
has been made in the direction of systematic proof engineering.

\section{Related Work}
\label{sec:related-work}

Numerous protocols have been designed to solve Byzantine agreement or BFT state-machine
replication~\cite{Castro2002, Kotla2007, Clement2009, El-Malek2005, Cowling2006,
cryptoeprint:2019:270, katz2006expected, abraham2019synchronous}, and the Abstract framework proposed by
\citet{Guerraoui2010} represents a notable step towards making the development of these protocols
more modular and systematic. The problem of Byzantine agreement has been known since 1980, when it
was first introduced by~\citet{Pease1980} as the ``Byzantine Generals Problem''. \citet{Dolev1982}
worked on signature-based protocols to solve this problem and introduced the principle of requiring
values to have more signatures in later rounds, which was also used by~\citet{Biely2007}.

To the best of our knowledge, there exists no full machine-checked proof of a state-of-the-art {BFT}
protocol. The most notable steps in this direction have been proofs of safety properties for {PBFT}
by \citet{Lamport2011} and more recently \citet{Rahli2018}. \citeauthor{Lamport2011} used the
{TLA\textsuperscript{+}} language in conjunction with the {TLAPS} proof system, while
\citeauthor{Rahli2018} developed an extension of {EventML} in the theorem prover {Coq} to verify the
protocol. The main issue with both of these proofs is that they omit any liveness properties, which
are commonly more challenging to prove than safety properties. We stress that these properties are
just as important to the security of the protocol, especially when deployed in a environment that
demands a reliable system. Another approach to verifying {PBFT} is based on {Event-B} and {Rodin},
but makes the assumption that messages cannot be forged~\cite{Krenicky2010}. A proof in this model
can therefore not provide any guarantees in the Byzantine fault model.

More success has been achieved for consensus algorithms that tolerate benign (i.e., non-Byzantine)
faults. In a breakthrough effort, the {IronFleet} project~\cite{Hawblitzel2015} fully verified a
{Paxos}-based distributed system and its implementation using the {Dafny} verifier. The
{Raft}~\cite{Ongaro2014} protocol is designed to achieve similar goals as the classic {Paxos}
algorithm, but with the main design goal of understandability. Many properties of the system and its
implementation have been formally verified in the {Verdi} framework for the {Coq} proof assistant,
but liveness properties were not
proved~\cite{Woos2016}.

In the only instance of a complete proof for a {BFT} protocol that we are aware of,
\citet{Debrat2012} verified two simple Byzantine consensus protocols in the Heard-Of model using the
{Isabelle/HOL} theorem prover. Our work builds on these results, showing that similar methods can be
applied to a more practical protocol that makes use of digital signatures.

Model checking is another method of verifying properties of protocols that can be used as an
alternative to formal proofs. A model checker explores the state space of a system specification,
which can be very useful for exposing design flaws without requiring a high amount of manual effort.
However, this technique is ineffective at providing assurance of security for a system due to the
constraints that combinatorial exploration places on the complexity of the model.
\citet{Zielinski2007} applied model checking to Byzantine consensus protocols, but due to the
tremendous size of the state space caused by Byzantine faults, the scope of the verification was
limited to only a single round of communication.

In order to address problems related to logging schemes such as Certificate
Transparency, \citet{Syta2015} proposed to leverage multisignatures and distribute
trust among many collective authorities or ``cothorities''. However, the rather
complex protocols used by cothorities are proposed without any formal verification.

\section{Conclusion}
\label{sec:conclusion}

This paper presents \protocol, a secure log replication protocol. To the best of our knowledge,
\protocol is the first practical BFT protocol to be accompanied by a complete machine-checked
security proof. This makes the protocol particularly suitable for security-critical applications
that demand high resilience and reliability.

We have demonstrated that our protocol can achieve sufficient throughput for practical applications
such as certificate logging, even in the presence of actively malicious participants. Concretely,
\protocol is able to support the load of the Certificate Transparency system for up to $25$ nodes.
Our system could be further improved by using a hash-tree based data store such as
Trillian~\cite{Eijdenberg2015}. A threshold signature scheme~\cite{Shoup2000} would also reduce the
size of the signed logs produced by the protocol.

Our extension of the Heard-Of model to signatures can be used in formal verification of other
consensus protocols that rely on this cryptographic primitive. Future work could capture other
primitives such as message authentication codes and hash functions. Other interesting paths for
future work include producing a verified implementation of \protocol and considering
application-specific optimizations.

\section{Acknowledgments}

We gratefully acknowledge support from ETH Zurich, and from the Zurich Information Security and Privacy Center (ZISC).
Furthermore, we thank Kartik Nayak, Dahlia Malkhi, and David Kozhaya for their valuable feedback.

\bibliographystyle{IEEEtranN}
\bibliography{paper.bib}

\begin{thebibliography}{43}
\providecommand{\natexlab}[1]{#1}
\providecommand{\url}[1]{#1}
\csname url@samestyle\endcsname
\providecommand{\newblock}{\relax}
\providecommand{\bibinfo}[2]{#2}
\providecommand{\BIBentrySTDinterwordspacing}{\spaceskip=0pt\relax}
\providecommand{\BIBentryALTinterwordstretchfactor}{4}
\providecommand{\BIBentryALTinterwordspacing}{\spaceskip=\fontdimen2\font plus
\BIBentryALTinterwordstretchfactor\fontdimen3\font minus
  \fontdimen4\font\relax}
\providecommand{\BIBforeignlanguage}[2]{{%
\expandafter\ifx\csname l@#1\endcsname\relax
\typeout{** WARNING: IEEEtranN.bst: No hyphenation pattern has been}%
\typeout{** loaded for the language `#1'. Using the pattern for}%
\typeout{** the default language instead.}%
\else
\language=\csname l@#1\endcsname
\fi
#2}}
\providecommand{\BIBdecl}{\relax}
\BIBdecl

\bibitem[Castro and Liskov(2002)]{Castro2002}
M.~Castro and B.~Liskov, ``Practical {Byzantine} fault tolerance and proactive
  recovery,'' \emph{{ACM} Transactions on Computer Systems}, vol.~20, no.~4,
  2002.

\bibitem[Kotla et~al.(2007)Kotla, Alvisi, Dahlin, Clement, and Wong]{Kotla2007}
R.~Kotla, L.~Alvisi, M.~Dahlin, A.~Clement, and E.~Wong, ``{Zyzzyva}:
  speculative {Byzantine} fault tolerance,'' in \emph{Proceedings of the {ACM}
  Symposium on Operating Systems Principles ({SOSP})}, 2007.

\bibitem[Clement et~al.(2009)Clement, Wong, Alvisi, Dahlin, and
  Marchetti]{Clement2009}
A.~Clement, E.~L. Wong, L.~Alvisi, M.~Dahlin, and M.~Marchetti, ``Making
  {Byzantine} fault tolerant systems tolerate {Byzantine} faults,'' in
  \emph{{NSDI}}, 2009.

\bibitem[Guerraoui et~al.(2010)Guerraoui, Kne{\v{z}}evi{\'{c}}, Qu{\'{e}}ma,
  and Vukoli{\'{c}}]{Guerraoui2010}
R.~Guerraoui, N.~Kne{\v{z}}evi{\'{c}}, V.~Qu{\'{e}}ma, and M.~Vukoli{\'{c}},
  ``The next 700 {BFT} protocols,'' in \emph{Proceedings of the European
  Conference on Computer Systems ({EuroSys})}, 2010.

\bibitem[Bessani et~al.(2014)Bessani, Sousa, and Alchieri]{Bessani2014}
A.~Bessani, J.~Sousa, and E.~E.~P. Alchieri, ``State machine replication for
  the masses with {BFT}-{SMART},'' in \emph{Proceedings of the {IEEE}/{IFIP}
  International Conference on Dependable Systems and Networks ({DSN})}, Jun.
  2014.

\bibitem[Lamport et~al.(1982)Lamport, Shostak, and Pease]{Lamport1982}
L.~Lamport, R.~Shostak, and M.~Pease, ``The {Byzantine} generals problem,''
  \emph{{ACM} Transactions on Programming Languages and Systems}, vol.~4,
  no.~3, Jul. 1982.

\bibitem[Zave(2012)]{Zave2012}
P.~Zave, ``Using lightweight modeling to understand {Chord},'' \emph{{ACM}
  {SIGCOMM} Computer Communication Review}, vol.~42, no.~2, Mar. 2012.

\bibitem[Woos et~al.(2016)Woos, Wilcox, Anton, Tatlock, Ernst, and
  Anderson]{Woos2016}
D.~Woos, J.~R. Wilcox, S.~Anton, Z.~Tatlock, M.~D. Ernst, and T.~Anderson,
  ``Planning for change in a formal verification of the {Raft} consensus
  protocol,'' in \emph{Proceedings of the {ACM} Conference on Certified
  Programs and Proofs ({CPP})}, 2016.

\bibitem[Lamport(2001)]{lamport2001paxos}
L.~Lamport, ``Paxos made simple,'' \emph{ACM Sigact News}, vol.~32, no.~4,
  2001.

\bibitem[Hawblitzel et~al.(2015)Hawblitzel, Howell, Kapritsos, Lorch, Parno,
  Roberts, Setty, and Zill]{Hawblitzel2015}
C.~Hawblitzel, J.~Howell, M.~Kapritsos, J.~R. Lorch, B.~Parno, M.~L. Roberts,
  S.~Setty, and B.~Zill, ``{IronFleet}: proving practical distributed systems
  correct,'' in \emph{Proceedings of the {ACM} Symposium on Operating Systems
  Principles ({SOSP})}, 2015.

\bibitem[Debrat and Merz(2012)]{Debrat2012}
\BIBentryALTinterwordspacing
H.~Debrat and S.~Merz, ``Verifying fault-tolerant distributed algorithms in the
  {Heard-Of} model,'' \emph{Archive of Formal Proofs}, 2012. [Online].
  Available: \url{https://isa-afp.org/entries/Heard_Of.html}
\BIBentrySTDinterwordspacing

\bibitem[Biely et~al.(2007)Biely, Charron-Bost, Gaillard, Hutle, Schiper, and
  Widder]{Biely2007}
M.~Biely, B.~Charron-Bost, A.~Gaillard, M.~Hutle, A.~Schiper, and J.~Widder,
  ``Tolerating corrupted communication,'' in \emph{Proceedings of the {ACM}
  Symposium on Principles of Distributed Computing}, 2007.

\bibitem[Nipkow et~al.(2002)Nipkow, Wenzel, and Paulson]{Nipkow2002}
T.~Nipkow, M.~Wenzel, and L.~C. Paulson, Eds., \emph{Isabelle/{HOL}}.\hskip 1em
  plus 0.5em minus 0.4em\relax Springer Berlin Heidelberg, 2002.

\bibitem[Laurie et~al.(2013)Laurie, Langley, and Kasper]{Laurie2013}
B.~Laurie, A.~Langley, and E.~Kasper, ``Certificate transparency,'' RFC 6962,
  Jun. 2013.

\bibitem[Basin et~al.(2014)Basin, Cremers, Kim, Perrig, Sasse, and
  Szalachowski]{Basin2014}
D.~Basin, C.~Cremers, T.~H.-J. Kim, A.~Perrig, R.~Sasse, and P.~Szalachowski,
  ``{ARPKI}: attack resilient public-key infrastructure,'' in \emph{Proceedings
  of the {ACM} Conference on Computer and Communications Security ({CCS})},
  2014.

\bibitem[Gritzalis(2002)]{Gritzalis2002}
D.~A. Gritzalis, ``Principles and requirements for a secure e-voting system,''
  \emph{Computers and Security}, vol.~21, no.~6, 2002.

\bibitem[Chondros et~al.(2015)Chondros, Zhang, Zacharias, Diamantopoulos,
  Maneas, Patsonakis, Delis, Kiayias, and Roussopoulos]{Chondros2015}
N.~Chondros, B.~Zhang, T.~Zacharias, P.~Diamantopoulos, S.~Maneas,
  C.~Patsonakis, A.~Delis, A.~Kiayias, and M.~Roussopoulos, ``A distributed,
  end-to-end verifiable, internet voting system,'' \emph{arXiv:1507.06812},
  2015.

\bibitem[Massias et~al.(1999)Massias, Avila, and Quisquater]{Massias1999}
H.~Massias, X.~S. Avila, and J.-J. Quisquater, ``Design of a secure
  timestamping service with minimal trust requirement,'' in \emph{Proceedings
  of the Symposium on Information Theory in the Benelux}, 1999.

\bibitem[Gipp et~al.(2015)Gipp, Meuschke, and Gernandt]{Gipp2015}
B.~Gipp, N.~Meuschke, and A.~Gernandt, ``Decentralized trusted timestamping
  using the crypto currency {Bitcoin},'' \emph{arXiv:1502.04015}, 2015.

\bibitem[Syta et~al.(2015)Syta, Tamas, Visher, Wolinsky, and Ford]{Syta2015}
E.~Syta, I.~Tamas, D.~Visher, D.~I. Wolinsky, and B.~Ford, ``Decentralizing
  authorities into scalable strongest-link cothorities,'' \emph{{CoRR}}, 2015.

\bibitem[pro()]{proofs}
\url{https://www.dropbox.com/s/w679b3xem11gjx4/proofs.tar.gz?dl=1},
  {Isabelle/HOL} source code.

\bibitem[Charron-Bost and Schiper(2009)]{Charron-Bost2009}
B.~Charron-Bost and A.~Schiper, ``The {Heard-Of} model: computing in
  distributed systems with benign faults,'' \emph{Distributed Computing}, Apr.
  2009.

\bibitem[Eijdenberg et~al.()Eijdenberg, Laurie, and Cutter]{VerifDataStruct}
A.~Eijdenberg, B.~Laurie, and A.~Cutter, ``Verifiable data structures,''
  \url{https://github.com/google/trillian/blob/master/docs/papers/VerifiableDataStructures.pdf}.

\bibitem[Dahlberg et~al.(2016)Dahlberg, Pulls, and
  Peeters]{dahlberg2016efficient}
R.~Dahlberg, T.~Pulls, and R.~Peeters, ``Efficient sparse {Merkle} trees,'' in
  \emph{Proceedings of the 21st Nordic Conference on Secure IT Systems
  (NordSec)}, 2016.

\bibitem[Chuat et~al.(2015)Chuat, Szalachowski, Perrig, Laurie, and
  Messeri]{gossip2015}
L.~Chuat, P.~Szalachowski, A.~Perrig, B.~Laurie, and E.~Messeri, ``Efficient
  gossip protocols for verifying the consistency of certificate logs,'' in
  \emph{Proceedings of the IEEE Conference on Communications and Network
  Security (CNS)}, September 2015.

\bibitem[Eskandarian et~al.(2017)Eskandarian, Messeri, Bonneau, and
  Boneh]{eskandarian2017certificate}
S.~Eskandarian, E.~Messeri, J.~Bonneau, and D.~Boneh, ``{Certificate
  Transparency} with privacy,'' in \emph{Proceedings on Privacy Enhancing
  Technologies}, 2017.

\bibitem[Li et~al.(2019)Li, Lin, Li, Wang, Li, Jing, and
  Wang]{li2019certificate}
B.~Li, J.~Lin, F.~Li, Q.~Wang, Q.~Li, J.~Jing, and C.~Wang, ``{Certificate
  Transparency} in the wild: Exploring the reliability of monitors,'' in
  \emph{Proceedings of the 2019 ACM SIGSAC Conference on Computer and
  Communications Security (CCS)}, 2019.

\bibitem[Pease et~al.(1980)Pease, Shostak, and Lamport]{Pease1980}
M.~Pease, R.~Shostak, and L.~Lamport, ``Reaching agreement in the presence of
  faults,'' \emph{Journal of the {ACM}}, vol.~27, no.~2, Apr. 1980.

\bibitem[Cloudflare()]{ct-stats}
Cloudflare, ``{Merkle Town},'' \url{https://ct.cloudflare.com}.

\bibitem[Klein et~al.(2009)Klein, Norrish, Sewell, Tuch, Winwood, Elphinstone,
  Heiser, Andronick, Cock, Derrin, Elkaduwe, Engelhardt, and
  Kolanski]{Klein2009}
G.~Klein, M.~Norrish, T.~Sewell, H.~Tuch, S.~Winwood, K.~Elphinstone,
  G.~Heiser, J.~Andronick, D.~Cock, P.~Derrin, D.~Elkaduwe, K.~Engelhardt, and
  R.~Kolanski, ``{seL}4: formal verification of an {OS} kernel,'' in
  \emph{Proceedings of the {ACM} {SIGOPS} symposium on Operating systems
  principles ({SOSP})}.\hskip 1em plus 0.5em minus 0.4em\relax {ACM} Press,
  2009.

\bibitem[Abd-El-Malek et~al.(2005)Abd-El-Malek, Ganger, Goodson, Reiter, and
  Wylie]{El-Malek2005}
M.~Abd-El-Malek, G.~R. Ganger, G.~R. Goodson, M.~K. Reiter, and J.~J. Wylie,
  ``Fault-scalable {Byzantine} fault-tolerant services,'' \emph{{ACM} Operating
  Systems Review}, vol.~39, no.~5, 2005.

\bibitem[Cowling et~al.(2006)Cowling, Myers, Liskov, Rodrigues, and
  Shrira]{Cowling2006}
J.~Cowling, D.~Myers, B.~Liskov, R.~Rodrigues, and L.~Shrira, ``{HQ}
  replication: a hybrid quorum protocol for {Byzantine} fault tolerance,'' in
  \emph{Proceedings of the Symposium on Operating Systems Design and
  Implementation ({OSDI})}, ser. OSDI '06, Berkeley, CA, USA, 2006.

\bibitem[Abraham et~al.(2019{\natexlab{a}})Abraham, Malkhi, Nayak, Ren, and
  Yin]{cryptoeprint:2019:270}
I.~Abraham, D.~Malkhi, K.~Nayak, L.~Ren, and M.~Yin, ``Sync hotstuff: Simple
  and practical synchronous state machine replication,'' Cryptology ePrint
  Archive, 2019.

\bibitem[Katz and Koo(2006)]{katz2006expected}
J.~Katz and C.-Y. Koo, ``On expected constant-round protocols for {Byzantine}
  agreement,'' in \emph{Annual International Cryptology Conference}, 2006.

\bibitem[Abraham et~al.(2019{\natexlab{b}})Abraham, Devadas, Dolev, Nayak, and
  Ren]{abraham2019synchronous}
I.~Abraham, S.~Devadas, D.~Dolev, K.~Nayak, and L.~Ren, ``Synchronous
  {Byzantine} agreement with expected {O(1)} rounds, expected {O($n^2$)}
  communication, and optimal resilience,'' in \emph{International Conference on
  Financial Cryptography and Data Security}, 2019.

\bibitem[Dolev et~al.(1982)Dolev, Fischer, Fowler, Lynch, and
  Strong]{Dolev1982}
D.~Dolev, M.~J. Fischer, R.~Fowler, N.~A. Lynch, and H.~R. Strong, ``An
  efficient algorithm for {Byzantine} agreement without authentication,'' in
  \emph{Information and Control}, 1982.

\bibitem[Lamport(2011)]{Lamport2011}
L.~Lamport, ``Byzantizing {Paxos} by refinement,'' in \emph{Lecture Notes in
  Computer Science}, 2011.

\bibitem[Rahli et~al.(2018)Rahli, Vukotic, Völp, and
  Esteves-Verissimo]{Rahli2018}
V.~Rahli, I.~Vukotic, M.~Völp, and P.~Esteves-Verissimo, ``{Velisarios}:
  byzantine fault-tolerant protocols powered by {Coq},'' in \emph{Programming
  Languages and Systems}, 2018.

\bibitem[Krenický and Ulbrich(2010)]{Krenicky2010}
R.~Krenický and M.~Ulbrich, ``Deductive verification of a {Byzantine}
  agreement protocol,'' Karlsruhe Institute of Technology, Department of
  Computer Science, Tech. Rep., 2010.

\bibitem[Ongaro and Ousterhout(2014)]{Ongaro2014}
D.~Ongaro and J.~K. Ousterhout, ``In search of an understandable consensus
  algorithm,'' in \emph{{USENIX} {ATC}}, 2014.

\bibitem[Zieli\'{n}ski(2007)]{Zielinski2007}
P.~Zieli\'{n}ski, ``Automatic verification and discovery of {Byzantine}
  consensus protocols,'' in \emph{Proceedings of the {IEEE}/{IFIP}
  International Conference on Dependable Systems and Networks ({DSN})}, Jun.
  2007.

\bibitem[Eijdenberg et~al.(2015)Eijdenberg, Laurie, and Cutter]{Eijdenberg2015}
A.~Eijdenberg, B.~Laurie, and A.~Cutter, ``Verifiable data structures,'' Google
  Research, Tech. Rep., 2015.

\bibitem[Shoup(2000)]{Shoup2000}
V.~Shoup, ``Practical threshold signatures,'' in \emph{Advances in Cryptology
  ({EUROCRYPT})}, 2000.

\end{thebibliography}

\appendix

\subsection{Protocol Formalization}
\label{sec:verification-protocol}

In the following, we show how \protocol can be specified within the extension of the Heard-Of model
presented in Section~\ref{sec:verification-sho_model} and provide formal specifications of the
security properties stated in Section~\ref{sec:problem}.

\def\X{\mathcal{X}}
\def\Votes{\mathcal{V}}
\def\Logs{\mathcal{L}}
There are two types of messages that are exchanged between nodes in \protocol:
\begin{enumerate}
    \item
    The first and most common one is used during the \phaseB phase and contains a list of value
    sets. Let $\X$ denote the domain of entries, and let $\Votes$ be the set of signatures over sets
    of values from $\X$, which can be modeled formally using the definition $\Votes \defeq \HOp
    \times (\HOp \times 2^\X)$ (consisting of the author, the identifier of the primary and the
    value to be signed). We call these signatures \emph{votes}. The set of possible messages of this
    type is then given by $M_1 \defeq \HOp \times 2^\Votes$, i.e., it carries a (possibly empty) set
    of votes for each primary node.

    \item
    During the \phaseC phase, each node sends out a single signature over a log. We refer to the
    domain of logs over entries from $\X$ as $\Logs_X$ and model the signatures over logs as $M_2
    \defeq \HOp \times \Logs_X$.
\end{enumerate}
These two message domains are combined in the set $M \defeq M_1 \union M_2$.

\def\Threads{\mathcal{T}}
Next, we specify the node states. Since every thread of the \algSub subroutine requires its own
state, we first define the possible \emph{thread states} as $\Threads \defeq 2^\X \times 2^{(2^\X)}
\times 2^\Votes$. This corresponds to the definitions in Algorithm~\ref{alg:sub}, which consist of
\begin{enuminline}
    \item the current decision $d \in 2^\X$,
    \item the sets of values $P \in 2^{(2^\X)}$ witnessed from the primary, and
    \item the votes $M' \in 2^\Votes$ from the current round.
\end{enuminline}
A node state stores such a thread state for each primary in $\HOp$, as well as the set of entries
collected from clients, the current version of the log and some signatures (from the set $\Sigma$)
for it:
\begin{align*}
    \HOst{i} \defeq \llpar~
        &\mathit{threads} \in \HOp \times \Threads,\\
        &\mathit{entries} \in 2^\X,\\
        &\mathit{log} \in \Logs_\X,\\
        &\mathit{sigs} \in 2^{M_2}
    ~\rrpar ~\text{for}~ i \in \HOp
\end{align*}
To simplify the notation, this is specified as a set of records, which are tuples whose components
are named. We use the syntax $s.\mathit{f}$ to access the field with name $f$ of a state $s$.

The initial states are defined as follows:
\begin{align*}
    \HOinit{i} \defeq \llpar~
        &\mathit{threads} = \HOp \times \{(\emptyset, \emptyset, \emptyset)\}),\\
        &\mathit{entries} = 2^\X,\\
        &\mathit{log} = \{L\},\\
        &\mathit{sigs} = \emptyset
    ~\rrpar ~\text{for}~ i \in \HOp
\end{align*}
where $L$ is some initial log that is the same for all nodes.

For the sake of brevity, we omit definitions such as the message sending functions and state
transition predicates. These are contained in the Isabelle theories, which are available
online~\cite{proofs} and were used in constructing the proofs.

Using the above definitions, it is now possible to formalize the security properties of the protocol
(defined in Section~\ref{sec:problem}):
\begin{LaTeXdescription}
        \item[Agreement] \emph{\propAgreement} \[
            \forall i, j \in \HOp \setminus \HObyz.~
            s_i^{f+2}.\mathit{log} = s_j^{f+2}.\mathit{log}
        \]
        Since the protocol consists of $f+2$ rounds, $s_i^{f+2}$ is the final state of the protocol
        run for a node $i$.

    \item[Completeness] \emph{\propCompleteness} \[
            \forall i \in \HOp \setminus \HObyz.~
            s_i^1.\mathit{entries} \subseteq s_i^{f+2}.\mathit{log}
        \]
            Clients and their requests are not modeled explicitly. Instead, each node uses the
            $\mathit{entries}$ field to store all received requests.  We can therefore express the
            property as follows: all requested entries that a node has received before the end of
            the \phaseA phase (i.e., $s_i^1.\mathit{entries}$) will be included in the log at the
            end of the protocol run.

    \item[Liveness] \emph{\propLiveness} \[
            \forall i \in \HOp \setminus \HObyz.~
            |s_i^{f+2}.\mathit{sigs}| \geq f+1
        \]
        A log is considered valid if it is signed by at least $f+1$ distinct nodes.
\end{LaTeXdescription}

\subsection{Proofs in Isabelle/HOL}
\label{sec:verification-isabelle}

\subsubsection{Using a Proof Assistant}

Proof assistants enable the construction of complex models and machine-checked proofs. Their
greatest advantage over other verifications tools is generality: they can be applied to
practically any system. This is done by constructing proofs that are analogous to handwritten
proofs, but provide much higher assurance since the proof assistant verifies its correctness,
checking that each step of the proof is sound. {Isabelle/HOL} is a particular instance of this type
of program, and its effectiveness has been proven in large-scale verification efforts such as that
of the {seL4} microkernel~\cite{Klein2009}.

In order to provide some intuition about how proofs are constructed with the assistance of
{Isabelle/HOL}, we show the code that is used to prove a simple statement about set theory, which is
also used during our verification of \protocol. It states that for any two sets $A$ and $B$ that are
subsets of a finite domain $C$, if the sum of their cardinalities is greater than that of the
domain, then $A$ and $B$ must overlap.

Isabelle can encode statements in higher-order logic (HOL), using the functional programming
language ML, which is enriched by a large variety of standard mathematical definitions. In
Isabelle's meta-language, the lemma stated above can be formalized as follows:

\begin{figure}[H]
    \input{example-lemma.tex}
\end{figure}

The keywords \isakeyword{assumes}, \isakeyword{and}, \isakeyword{shows} are part of Isabelle's
meta-language. They identify the premises and conclusions of the lemma, which can be named as in
this example to improve readability of the code. Isabelle supports reasoning using declarative
proofs in its proof language Isar, which closely resembles the structure of handwritten proofs.
Moreover, Isabelle provides a rich library of common lemmas that can be used in proofs.

For this lemma, we can use a proof by contradiction. First, we assume that $A \cap B = \{\}$, from
which we can deduce two contradicting results:
\begin{itemize}
    \item
    $|A \union B| > |C|$.
    Given that $A$ and $B$ do not intersect, $|A| + |B| = |A \union B|$ holds. This follows from
    the built-in rule $\mathit{card{\isacharunderscore}Un{\isacharunderscore}disjoint}$, which is
    not defined here for the sake of brevity.
    \item
    $|A \union B| \leq |C|$.
    This follows from the fact that since $A$ and $B$ are both subsets of $C$, it holds that
    $A \union B \subseteq C$. The built-in rule $\mathit{card{\isacharunderscore}mono}$ is used to
    convert this into the equation about cardinalities.
\end{itemize}
From these two statements, we can show $\mathit{False}$, which is a contradiction.
The code for the complete proof in Isabelle is shown below:

\begin{figure}[H]
    \input{example-proof.tex}
\end{figure}

The detailed syntax and semantics of the proof are out of scope for this brief overview of Isabelle,
but thanks to the structure that resembles the language of handwritten proofs, the steps are fairly
comprehensible. For each step of the proof, an automated proof strategy (such as $\mathit{auto}$,
$\mathit{simp}$, \ldots) is given using a \isakeyword{by} keyword, providing a hint to the proof
assistant about how the step can be automatically verified. Once the machine is able to verify all
individual steps, the lemma has been proved.

This particular lemma was chosen as an example due to its simplicity, but is also a part of our
proofs, where we use it to show that a set of $f+1$ nodes always contains at least one correct node.
Our Isabelle theory contains a total of $104$ lemmas, most of which are more complex to express and
prove than the one shown above. However, this example serves as an overview of what to look for when
performing high-level inspection of the proofs.

\subsubsection{Proof Methodology}

In the following, we provide a high-level description of the process of constructing our proofs for
\protocol. Initially, we formalized the protocol specification (given by pseudocode) in the HO model
and specified the desired properties formally. We have found it an effective strategy for the
construction of a sophisticated proof to proceed along the following basic steps:
\begin{enumerate}
    \item Write down a sequence of informal arguments for why the properties hold (similar to our
        analysis in Section~\ref{sec:analysis-security}). The goal of this step is to identify key
        intermediate lemmas around which the proof will be constructed.
    \item Formalize the intermediate lemmas from the previous step to obtain a rough outline in the
        proof assistant. Insert placeholders where the proof is incomplete (e.g., using the
        \isakeyword{sorry} keyword in Isabelle) to enable the assistant to check this outline.
    \item Attempt to complete the unfinished proof steps, using existing lemmas and creating new
        ones where required. Repeat this step until all statements are proved.
\end{enumerate}
This strategy is similar to a backward search technique, where the goal is inspected first and
lemmas are created whenever they are needed to solve open goals.

Naturally, this description of the process is fairly idealized, as it is often necessary in practice
to modify the original outline or even the protocol itself upon discovering that certain steps of
the proof cannot be completed as expected. This is a common occurrence, as it is extremely difficult
for a human to anticipate the precise outline of a proof before carrying out the steps in detail.

It is therefore necessary to ensure that the proofs are robust to changes and highly automated, such
that a small change in some part of the proof does not require rewriting all of the remaining steps.
For this purpose, we employ the design concept of modularization, which is standard folklore in the
software engineering domain thanks to its benefits in creating maintainable code. In the case of a
formal proof, this concept implies that the interfaces at which certain components of the model
interact must be specified clearly and kept as abstract as possible. Isabelle provides useful
concepts for this, called \emph{locales} and \emph{definitions}, which fulfill a similar role as
abstract interfaces in software engineering. Well-designed modular proofs enable changing details of
the protocol or lemmas without completely rewriting the proof.

An example of how our proof utilizes this concept is the \algSub subroutine, which is invoked in
parallel by \protocol. It is natural to first prove the properties of the subroutine, providing an
interface from which a proof for the properties for \protocol to be constructed. While this
composition appears to be simple, we discovered a possible attack where it was possible for an
adversary to replay legitimate signatures across different threads, preventing the correct nodes
from creating valid logs.

\end{document}